\documentclass[preprint]{aastex}

\def\l{{\em l}}

\def\d{{\em d }}

\shorttitle{Modeling Fe K Line in AGN with a Disk Wind}
\shortauthors{Tatum et al.}

\begin{document}

\title{Modeling the Fe K Line Profiles in Type I AGN with a Compton-Thick Disk Wind}
\author{M.M. Tatum\altaffilmark{1}, T.J. Turner\altaffilmark{1}, S.A. Sim\altaffilmark{2}, 
L. Miller\altaffilmark{3}, J.N. Reeves\altaffilmark{4},  A.R. Patrick\altaffilmark{4}, K.S. Long\altaffilmark{5}}

\altaffiltext{1}{Department of Physics, University of Maryland Baltimore County, Baltimore, MD 21250}
\altaffiltext{2}{Research School of Astronomy and Astrophysics, Mount Stromlo Observatory, Cotter Road, Weston Creek, ACT 2611, Australia}
\altaffiltext{3}{Dept. of Physics, University of Oxford, 
Denys Wilkinson Building, Keble Road, Oxford OX1 3RH, U.K.}
\altaffiltext{4}{Astrophysics Group, School of Physical and Geographical Sciences, Keele 
University, Keele, Staffordshire ST5 5BG, U.K}
\altaffiltext{5}{Space Telescope Science Institute, 3700 San Martin Drive, Baltimore, MD 21218, U.S.A}

\begin{abstract}

We have modeled a small sample of Seyfert galaxies that were previously identified as having simple X-ray spectra with little intrinsic absorption. The sources in this sample all contain moderately broad components of Fe K-shell emission and are ideal candidates for testing the applicability of  a Compton-thick accretion-disk wind model to AGN emission components. Viewing angles through the wind allow the observer to see the absorption signature of the gas, whereas face-on viewing angles allow the observer to see the scattered light from the wind. We find that the Fe K emission line profiles are well described with a model of a Compton-thick accretion-disk wind of solar abundances, arising tens to hundred of gravitational radii from the central black hole. Further, the fits require a  neutral component of Fe K$\alpha$ emission that is too narrow to arise from the inner part of the wind, and likely comes from a more distant reprocessing region. Our study demonstrates that a Compton-thick wind can have a profound effect on the observed X-ray spectrum of an AGN, even when the system is not viewed through the flow. 

\end{abstract}


\section{Introduction}
Active galactic nuclei (AGN) have rich spectra with a multitude of absorption and emission signatures, which offer insight into the  structure of the nuclear environs. Current X-ray observatories, {\it Suzaku}, {\it Chandra}, and {\it XMM-Newton}, can resolve H- and He-like species of C, N, O, Ne, Mg, Al, Si, and S (e.g., \citealt{Brandt:2002kx}).  Absorption lines from these species generally show blueshifts \citep[typically in the range 100 $\sim$ 1000 km s$^{-1}$,][]{Blustin:2005mz}, indicative of an origin in outflowing gas. 

In addition, features from a range of ionization states of Fe have been resolved including the neutral Fe K$\alpha$ emission line at $\sim$6.4 keV; the H-like  and He-like  species at higher energies, which are often seen in emission and absorption;  a broad  unresolved transition array from the M-shell just below 1 keV (e.g. \citealt{Behar:2001ys});  and Fe L transitions, which are usually seen in absorption and sometimes weakly in emission (e.g. \citealt{Kaspi:2002lr}). Observations of a strong Fe L emission signature are sometimes evident in AGN spectra but generally attributed to starburst activity, e.g. NGC 1365 \citep{Guainazzi:2009ly} and  Mrk 573 \citep{Bianchi:2010gf}.

In principle, the profile of Fe K$\alpha$ emission can be used to constrain conditions in the reprocessing gas. The line is sufficiently strong that useful constraints may be obtained from currently-available data. However, interpretation of the line profile  has been challenging because of the difficulty in separating line and continuum photons in X-ray data. Some early work suggested the line to be extremely broad, owing to strong relativistic effects (e.g. \citealt{Reynolds:1995lr} and \citealt{Fabian:2002fk}). Such an interpretation required the line be produced within a few r$_g$ of the black hole. An {\it ASCA} sample study suggested relativistically-blurred lines exist in $\sim$ 75\% of local Seyfert galaxies \citep{Nandra:1997mz}, although later work  using {\it XMM} data \citep{Nandra:2007lr} concluded that only  $\sim$ 40\% of local Seyferts showed evidence for  lines originating within 50r$_g$. Subsequently, analysis of a  flux-limited sample  \citep{de-La-Calle-Perez:2010qy}, reduced the estimate of extremely broad Fe K$\alpha$ emission to $\sim$ 30\% of sources.  However, strong evidence for high column X-ray absorbers in the local AGN population has motivated several teams to reconsider the true line profile and origins of the Fe K-shell emission, and it may be that all claimed relativistically-blurred lines are actually a manifestation of absorption and scattering through circumnuclear gas \cite[e.g.][]{Miller:2008kx}.

In a study of NGC 3783, \citet{Reeves:2004qy} found that modeling all of the absorbing zones accounted for much of the curvature originally attributed to the broadness of the Fe K emission line. In NGC 4051, detailed modeling of the absorber complex led to a limit  FWHM $< $ 5540 km s$^{-1}$ for the moderately broad component of Fe K emission \citep[e.g.][]{Turner:2010kx}. In contrast, when partial-covering absorption was not accounted for in model fits, the mean FWHM line width suggested by \citet{Nandra:2007lr} was $\sim$ 37000 km s$^{-1}$ for a sample of Seyfert 1 galaxies.  When the absorbing complex is fully accounted for there remains no requirement for a relativistically-blurred component of Fe K$\alpha$ emission in any source, not even MCG-6-30-15 \citep{Miller:2008kx}. However, a very modest broad component is sometimes evident that does not require the invocation of relativistic effects, and this is the subject of our paper. 

To explain the grating spectra of the best studied sources (e.g. MCG-6-30-15, \citealt{Lee:2001fk}, \citealt{Miller:2008kx}), multiple zones of ionized, absorbing gas are required.  Of particular interest is the detection of strong absorption from Fe {\sc xxv} and Fe {\sc xxvi}, requiring column densities $>$ 10$^{23}$ cm$^{-2}$ in the line-of-sight for some sources \citep[e.g.][]{Pounds:2003uq,Reeves:2003uq,Risaliti:2005uq, Miller:2007fe, Turner:2008qy}. Such high-column absorbing systems  appear to be common in radio-quiet AGN, being detected in more than 40\% of sources studied by \citet{Tombesi:2010yq} , and typically showing significant outflow velocities from a few thousand km s$^{-1}$ upwards to $\sim$0.3\,c. 

The strong evidence for Compton-thick outflows in type 1 AGN motivated development of a multi-dimensional radiative-transfer realization of a Compton-thick wind (e.g. \citealt{Sim:2005uq,Sim:2008ys,Sim:2010fr}). Such winds are expected in AGN having a high Eddington ratio \citep{King:2003fk} and these would likely be launched fairly close to the black hole \citep[e.g.][]{Proga:2000wd}. Consideration of high wind densities led to the prediction of strong wind signatures, including deep absorption lines and a moderately broadened Fe K-shell emission profile. 

\citet{Sim:2008ys,Sim:2010fr} have calculated wind spectra assuming a launched radius beyond 30 gravitational radii, such that relativistic blurring does not significantly affect the line width. These simulations show a moderately broadened line profile having FWHM $\sim$ 1 keV owing to the combined effects of Compton scattering and the large rotational/outflowing velocities in the wind. Electron scattering in the flow combined with photoelectric absorption contributes to a slightly red-skewed wing to the observed Fe K emission profile (c.f. \citealt{Laurent:2007fj}). The Fe K$\alpha$ line is a strong signature in the X-ray spectrum of the wind, with equivalent widths in excess of  $\sim 200$ eV for some wind parameters \citep{Sim:2008ys}.  

Thus, a Compton-thick reprocessing wind simultaneously produces both blue-shifted absorption lines and red-skewed emission lines in the Fe K regime for AGN, the relative strength of these in the observed spectrum depends upon the orientation of the flow with respect to the observer's line-of-sight. 

In this paper, we focus on the interpretation of observable features in the Fe K band in the context of a Compton-thick wind.  Although other spectral lines are expected to form  in the flow, identification and isolation of features in the Fe K energy band is more secure than in the soft X-ray regime, since only K shell transitions of heavy ions are expected to be present around 6 - 7 keV. Although  Fe K-shell absorption features are excellent diagnostics for a highly ionized flow, the associated Fe K emission arising from line scattering or recombination in the outflow is also important and has only been assessed for two AGN to date (see \citealt{Sim:2010fr}). Specifically,  case studies of  Mrk 766 and PG1211$+$143 have shown that  absorption and emission features in the  Fe-K  regime can successfully be explained using a single Compton-thick flow (see \citealt{Sim:2010fr} and references therein).   

The best constraints on the emission line profile can be obtained when there are no confusing absorption features present; these are difficult to resolve spectroscopically using current data. With this in mind, the small sample identified by \citet{Patrick:2011mz} is of particular interest; that sample comprises objects showing little evidence of intrinsic X-ray absorption.  Unobscured objects present a relatively clear view of the Fe K region enabling a fit to the broadband spectrum and a basic understanding of the Fe K emission features that is unencumbered by consideration of separating the signature of a complex X-ray absorber. 

\citet{Patrick:2011mz} found a moderately broad component of Fe K$\alpha$ to exist in all of their sample sources. Here we test whether these moderately broad Fe-K emission, together with limits on absorption signatures, components are consistent with an origin arising from reflection off a Compton-thick wind. We begin in Section 2 by outlining the parameters of the wind model. In Section 3, we  outline the observations and data reduction; and in Section 4, we detail the spectral fitting process and results. Section 5 discusses the implications of our results. We draw conclusions in Section 6. 

\section{The Compton-thick Wind Model}

To model the Compton-thick wind, we used a model grid based on the Compton-thick wind model of \citet{Sim:2010fr}. We present a general overview of the Compton-thick wind model and refer the reader to the aforementioned paper for specific details.  

\citet{Sim:2010fr} presented the results of Monte Carlo radiative transfer calculations, simulating a Compton-thick disk wind. The wind was parameterized as a steady-state, continuous,
bi-conical outflow. The density in the flow varied smoothly as a function of position and was determined by the geometry of the flow, the total mass-loss rate (\.{M}$_w$) and the relative mass-loading of streamlines in the flow. Hydrodynamical simulations \citep[e.g.][]{Proga:2004vn} showed that disk winds were expected to have very complex, non-spherically symmetric geometries. However, \citet{Sim:2010fr} invoked a simple velocity parameterization because it allowed for a relatively simple geometry. The model does not claim to offer a completely realistic description of the flow, but approximates the key physical properties of the wind and makes it possible to understand the influence of these parameters on the observed spectrum, by computing model grids (see \citealt{Sim:2008ys,Sim:2010fr}). Comparison between this version of the model and points from  a more complex, hydrodynamical model \citep{Sim:2010zr}  suggest that the velocity prescription can yield predictions for the wind that are close to those obtained with a much more time-consuming, intensive calculation. 

The geometry adopted a modification of the ``displaced dipole'' model described in \citet{Knigge:1995qf}.  Three parameters defined the extent and opening angle of the wind: \d was the distance below the origin to which all the stream lines in the wind converge; {\em R$_{min}$} was distance from the origin to the innermost edge of the wind in the xy-plane in units of  gravitational radius, r$_g$; and {\em R$_{max}$} was the distance from the origin to the outermost edge of the wind in the xy-plane (see Figure 1 in \citealt{Sim:2008ys}). The disk was assumed to lie in the xy-plane and the wind was symmetric under rotation about the z-axis and under reflection in the xy-plane. 

\citet{Sim:2010fr} invoked a prescription that specifies the velocity at every point in the wind following the parameterization of \citet{Knigge:1995qf}. It was assumed that parcels of matter conserve specific angular momentum about the z-axis as they flowed outwards with the angular momentum at the base of a streamline set at the Keplerian value for the radius at which this rose from the xy-plane. Thus, the rotational velocity was determined  by the specification of wind geometry and the mass of the central black hole. The poloidial velocity equation was \begin{displaymath} v_l=v_0 +(v_\infty - v_0)(1-\frac{R_v}{R_v + l})^\beta\end{displaymath} where \begin{math}v_\infty=f_v\sqrt{\frac{2GM_{bh}}{R}}\end{math} was a user specified multiple, \begin{math} f_v\end{math}, of the escape velocity from the base of the streamline in the xy-plane, \begin{math} M_{bh}\end{math} was the black hole mass, \begin{math} v_0\end{math} was the initial velocity,  {\em l} was the distance along the outflow streamline, $R_v$ was an acceleration  length parameter, and $\beta$ set the rate of acceleration. Varying the $\beta$, R$_v$ and f$_v$ parameters allowed us to explore the influence of different rates of wind acceleration in the model. Ultimately, this related to the physical process responsible for driving the flow.

As discussed by \citet{Sim:2008ys}, the Monte Carlo simulations relied on discrete, indivisible packets of radiation based on \citet{Lucy:2002tw}. In order to calculate the radiative transfer, a 100 x 100 wind  grid was produced where the wind properties was discretized in terms of the parameters $R_v$ and \l. This grid was transformed into a 3D 100 x 100 x 100 Cartesian grid, where  those grid points inside the wind were assigned density, ionization/excitation states and radiation field properties of the point closest to the wind  grid point. Those grid points outside the wind remained empty. These discrete packets ranged from 0.1 keV to 511 keV with an assumed power-law distribution with the photon index ($\Gamma$) as an input parameter. The X-ray source was assumed to be a spherical shell with radius 6 r$_g$, so that the X-ray emitting source had a non-zero angular size. The packets propagated until they reached the outer boundary of the computational domain. They may have interacted with the wind via Compton scattering, bound-bound, bound-free (including photoelectric absorption by inner shell electrons) and free-free processes. Following interactions, the Macro Atom method of \citet{Lucy:2002tw, Lucy:2003os} was used to simulate subsequent fluorescent or thermal re-emission by the outflow. During each Monte Carlo simulation, the wind properties (e.g. ionization state, temperature) were held fixed. Once all packets propagated through the outflow, the packet histories were used to compute photo-ionization and heating rates in each grid cell. These were then used to solve the equation of ionization and thermal equilibrium and determine improved estimates for the ionization state and temperature. These estimates were then used as input to a new Monte Carlo simulation. Thus, via a sequence of Monte Carlo simulations, we iteratively determined the conditions in the outflow. The total output spectra contained two components: a direct photon spectrum, containing photons that had no interaction with the wind, and a scattered/reprocessed photon spectrum, containing photons that interacted with the wind. The output spectrum was not calculated until the last iteration of the simulation.

\subsection{Construction of synthetic wind spectra for spectral fitting}   
To generate a set of synthetic wind spectra to use for fitting the data, we performed radiative transfer calculations for a grid of models. In the grid, we varied the following parameters of the model: \.{M}$_w$, $d$, R$_{min}$, R$_{max}$, Fe abundance and $\Gamma$. The values used on the grid are listed in Table~\ref{tab:syn}.  

The spectral signature from the  Compton-thick wind model of \cite{Sim:2010fr} was calculated over a range of parameter-space. We assumed  $f_v=1$, which is consistent with a radiatively driven flow. The parameters $R_v$ and $\beta$  depend on the physical acceleration process and the wind location; we assumed  $R_v=R_{max}$ in all cases and $\beta=1$, and these represent acceleration which occurs on the scale of the system. We assumed $v_0 \ll v_{\infty}$\,, and therefore adopted $v_0=0$\,. A black hole mass M\,$=10^7$ M$_\odot$ was utilized, with an intrinsic 2-10 keV luminosity L$_X=0.01$ L$_{Edd}$, where L$_{Edd}$ was the Eddington luminosity, and a mass-loss exponent $\kappa$= --1, which was chosen because this gives a gentle density variation with radius making the ionization structure of the wind easier to resolve. The outer radius of the simulation grid, i.e. the extent of the region over which the influence of the wind on the spectrum was calculated, was 5 x 10$^{16}$ cm. For models with higher Fe abundance, we also increased the Ni abundance by the same factor. 

Given the limited spectral resolution and signal-to- noise ratio of the data, we found that it was not possible to determine all wind parameters simultaneously. As a result of these parameter degeneracies, we fixed both the inner radius and the focal point {\em d}  at 32 r$_g$; the outer radius to $\sim$ 300 r$_g$ and the Fe abundance to solar abundance, via \citet{Asplund:2005lr}. This geometry fixed the opening angle to $\sim$ 49$^o$. We further discuss the parameter degeneracies in Section 5. Although a black hole mass is specified, the simulated wind is applicable to other black hole mass values as all parameters scale with black hole mass. This particular black hole mass was chosen as a convenient scaling factor.  

For each model on the grid, we computed synthetic spectra for 20 observer orientations. We therefore generated a total of 4320 synthetic spectra.
\section{Observations}
{\em Suzaku} has four X-ray telescopes, each containing a silicon CCD within its focal plane forming the X-ray Imaging Spectrometers (XIS) suite. XIS0, XIS2 and XIS3 are front-illuminated (FI), providing data over a usable range of 0.6-10.0 keV with an energy resolution   FWHM  $\sim$ 130 keV at 6.0 keV. In November 2006, a charge leak was discovered in XIS2, making XIS0 and XIS3 the only operational FI chips. XIS1 is back-illuminated. The back-illuminated configuration extends the soft band to $\sim$0.2 keV; however, it also results in a lower effective area and higher background rate in the Fe K regime, compared to the FI chips. Consequently, XIS1 is excluded from our spectral analysis. Suzaku also carries the Hard X-ray Detector (HXD) that contains a silicon PIN diode detector covering a range of 10.0--100 keV with a usable energy range of 15.0--70.0 keV. 

The {\em Suzaku} observations were made 2006 August 28 (OBSID 701031010), 2007 April 1 (OBSID 702014010), 2007 June 7 and 2010 May 19 (OBSID 702043010, 705063010), 2007 December 9 (OBSID 702122010), 2008 June 24 (OBSID 703028010), and 2008 August 28 (OBSID 703060010) for Mrk 335, Ark 120, Fairall 9, SWIFT J2127.4+5654, NGC 7469, and MCG-02-14-009, respectively. An observation log, observation exposure times and source count rates are detailed in Table 2 of \citet{Patrick:2011mz}.   

The data were reduced using HEAsoft v.6.10. The XIS cleaned event files were screened in XSELECT to exclude data  during passage through the South Atlantic Anomaly and also  excluding data starting 500 s before entry  and up to 500 s after exit. In addition, we excluded data having  an Earth elevation angle $<$ 10$^{\circ}$ and a cut-off rigidity $>$ 6 GeV.  CCDs were in 3 x 3 and 5 x 5 edit modes, with normal clocking mode. Good events were selected, having  grades 0, 2, 3, 4, and 6, while  hot and flickering pixels were removed using the SISCLEAN script. The spaced-row charge injection was utilized. XIS spectra were extracted from circular regions of 3.0' radius centered on the source, while the background was extracted from a region of the same size offset from the source and from the corners of the chip that register calibration data. 

Data from the front-illuminated CCDs XIS0 and 3 were used for all observations, and data from the two chips were were co-added using {\sc addascaspec}.  For Mrk~335, data were also available from XIS2, and these were combined with XIS0,3 in that case. For  SWIFT J2127.4+5654,  only  data from the XIS3 chip were used, since the XIS0 chip was not functioning properly at the time of observation. 

The cleaned PIN data was reduced utilizing the ftool {\sc hxdpinxbpi}. This tool calculates good time intervals (GTIs) of the non X-ray instrumental background (NXB, using model 'D' released 2008 June 17\footnote{http://www.astro.isas.jaxa.jp/suzaku/doc/suzakumemo/suzakumemo-2007-01.pdf} ) data that overlap with the source data and extracts both the source spectrum and NXB background spectrum through that common GTI. A simulated cosmic X-ray background (CXB) spectrum is calculated and combined with the NXB spectrum to produce a total PIN background spectrum for the observation. A dead-time correction $\sim$ 4-5\% is  applied to the source spectrum using the ftool {\sc hxddtcor}. Finally, the PIN data have a known 1$\sigma$ systematic uncertainty of 1.3\%\footnote{http://heasarc.nasa.gov/docs/suzaku/analysis/watchout.html}, which was applied to the PIN data in {\sc grppha}.  
\section{Spectral Fitting Results} 

Spectral analysis was performed using {\sc  XSPEC v 12.5}. To fit these data, the 4320 synthetic spectra previously mentioned were used to create a multiplicative model table (mtable). To adjust for instrumental calibrations during spectral fitting, the PIN model  was scaled by an energy-independent constant as appropriate for the HXD and XIS nominal aim points, respectively. The constant value was  1.18 for all observations except Fairall 9 (OBSID 702043010), for which a constant scalar of 1.16 was appropriate. Prior to fitting, the XIS data were binned to the half-width half maximum resolution of the instrument, where as the PIN data were binned to $\Delta$log$_{10}$ Energy= 0.1 resolution. In the fitting process, the $\chi^2$ statistic was minimized and then errors were calculated for the 90\% confidence interval for each parameter. 

First, we fit  the 2-10 keV band, to investigate whether the wind model can account for the Fe K-shell spectral signature in the sample sources. We note that the two observations of Fairall 9 were fit simultaneously to constrain parameters that should not differ between epoch, such as viewing angle and Galactic column. The spectral model was constructed using a powerlaw continuum of free normalization and photon index $\Gamma$. The continuum form was modified by the absorption and remission from the Compton-thick wind and the illuminating spectrum for the wind was assumed to be the  power-law continuum. As stated previously, the degeneracies in the model have led us to allow for only three free parameters in the wind model: $\Gamma$, \.{M}$_w$ and $\theta$.
 
\citet{Patrick:2011mz} previously identified several components of Fe K$\alpha$ emission in the sample sources: most objects show clear moderately broad and narrow components of Fe K$\alpha$ emission. To account for the core of the Fe emission line, which is generally  too narrow to be  produced in the wind, we included a narrow ($\sigma$= 10 eV) Gaussian model component. All models also included a Galactic component of absorption parameterized using the {\sc tbabs} model \citep{Wilms:2000qy} and using values from the weighted average $N_H$ in the Dickey \& Lockman HI in the Galaxy survey \citep{Dickey:1990uq}. 

\begin{figure}
\rotatebox{-90}{\includegraphics[width=5cm]{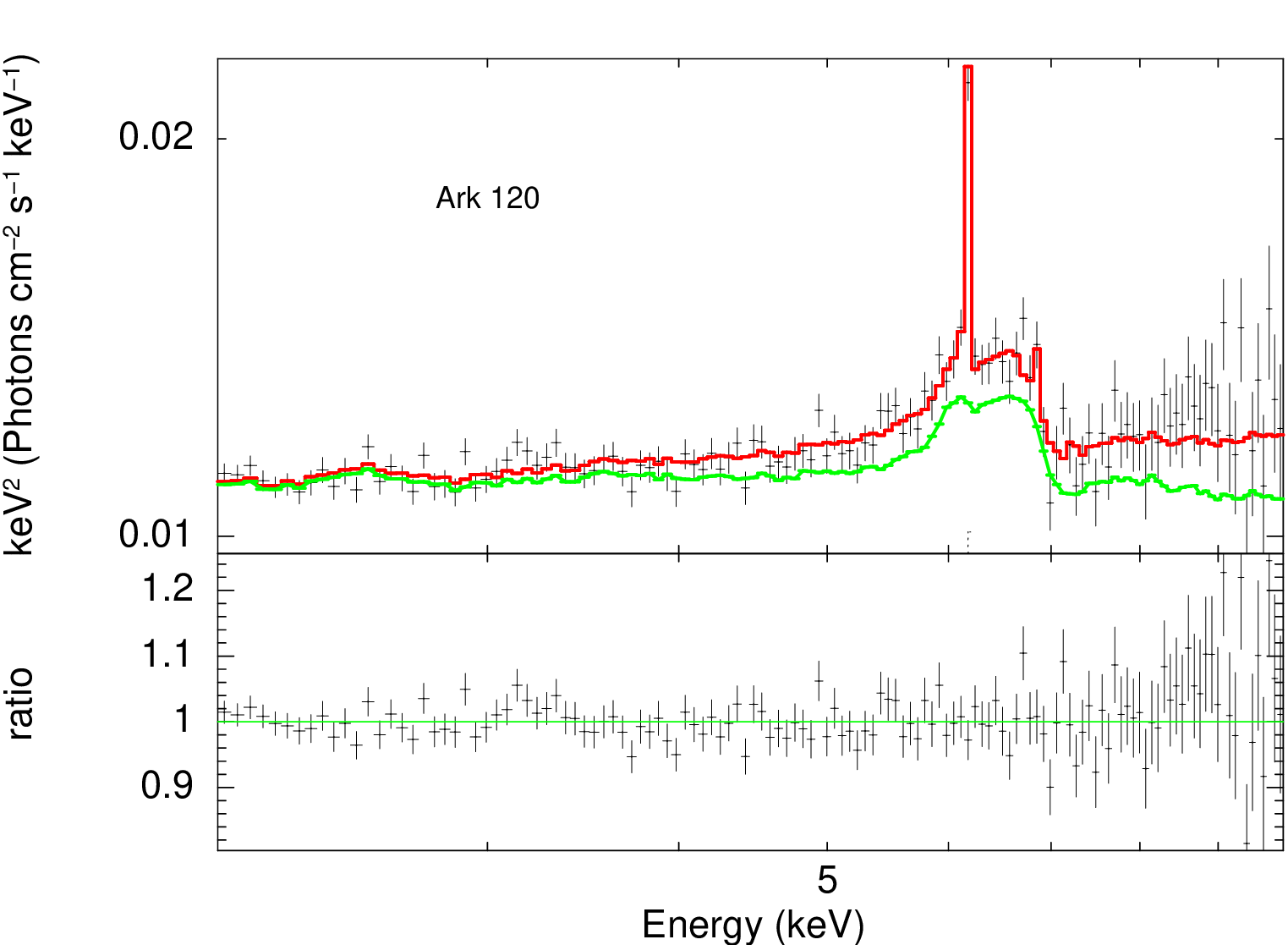}}
\rotatebox{-90}{\includegraphics[width=5cm]{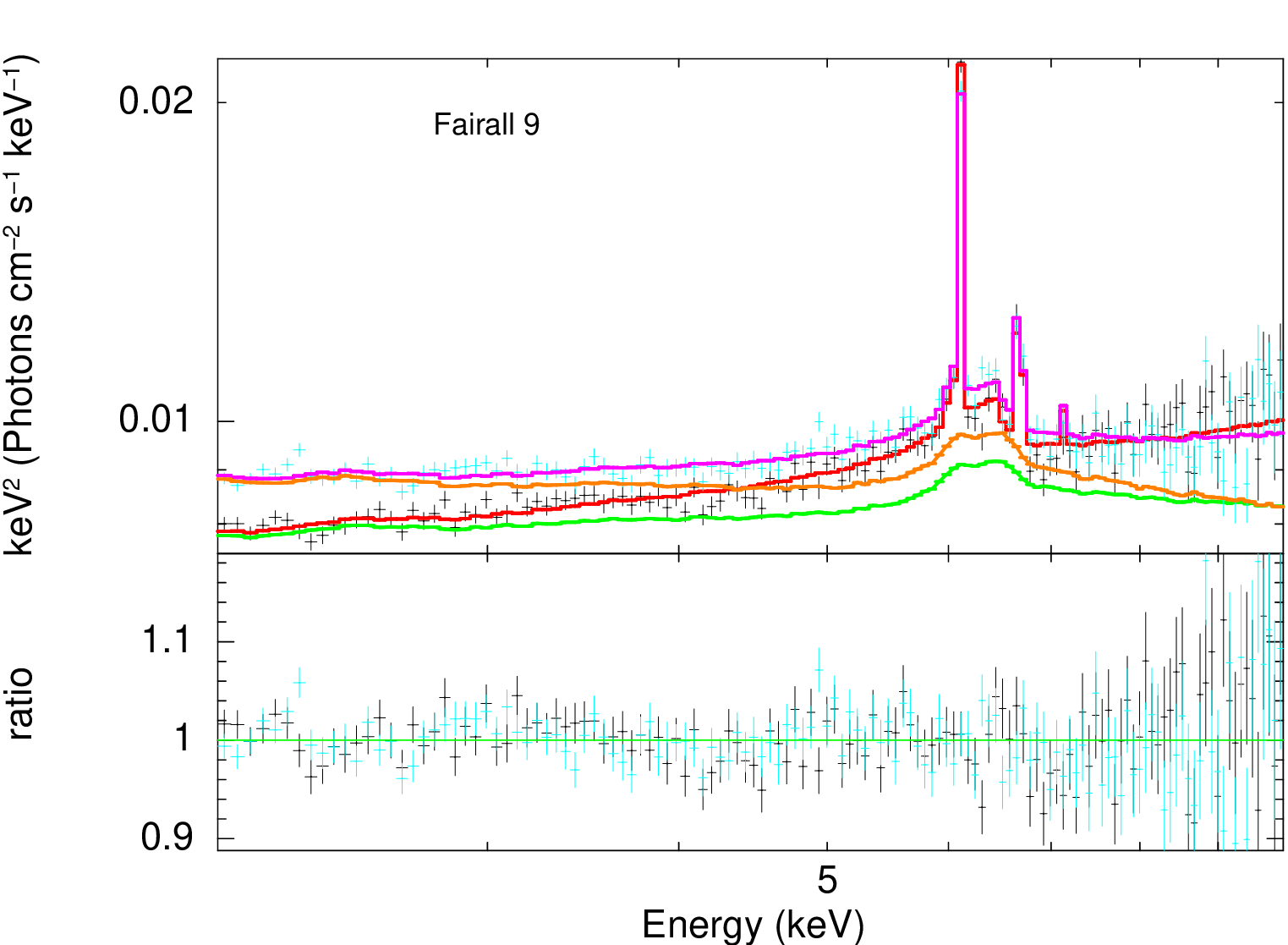}}
\rotatebox{-90}{\includegraphics[width=5cm]{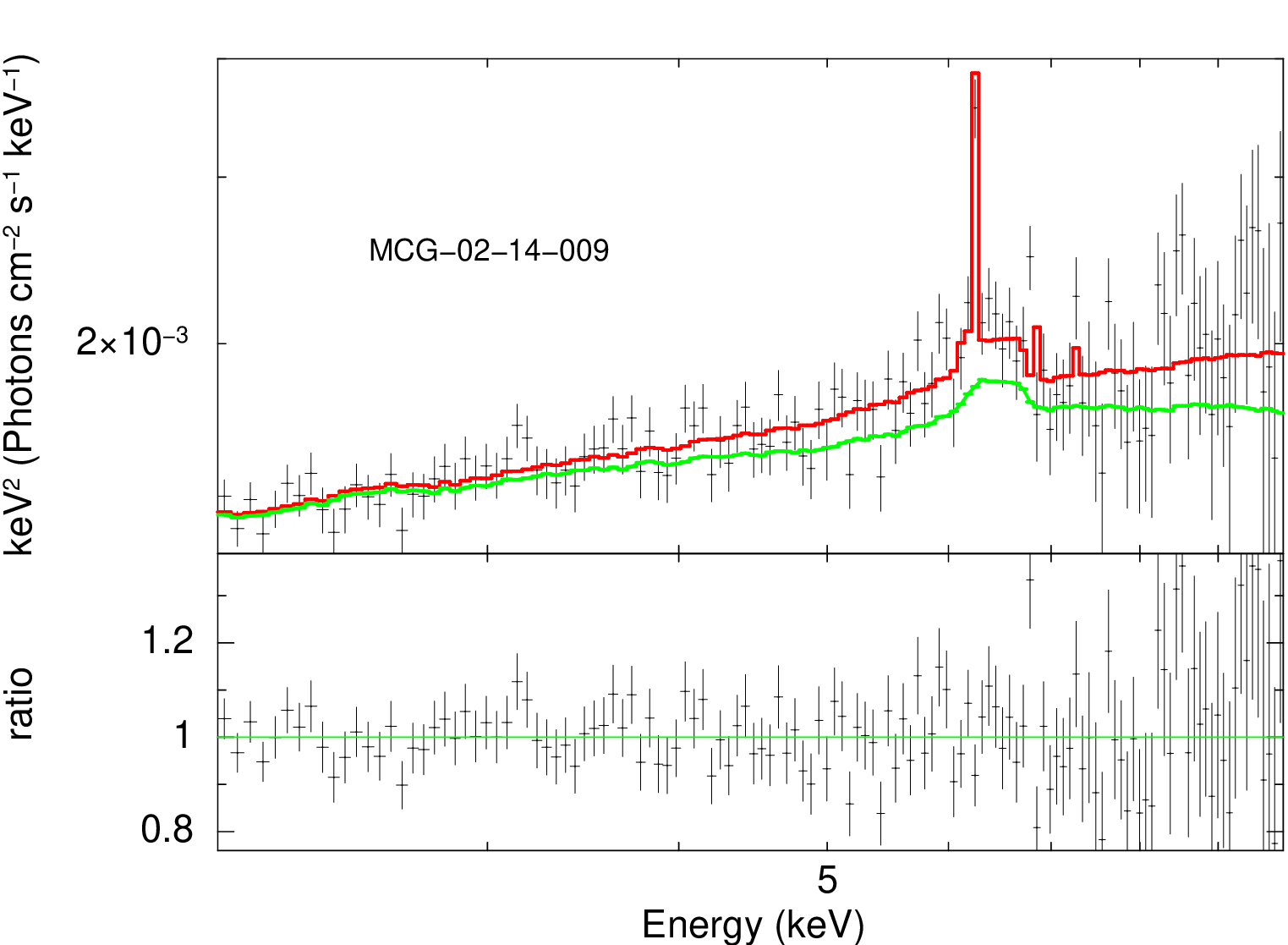}}
\rotatebox{-90}{\includegraphics[width=5cm]{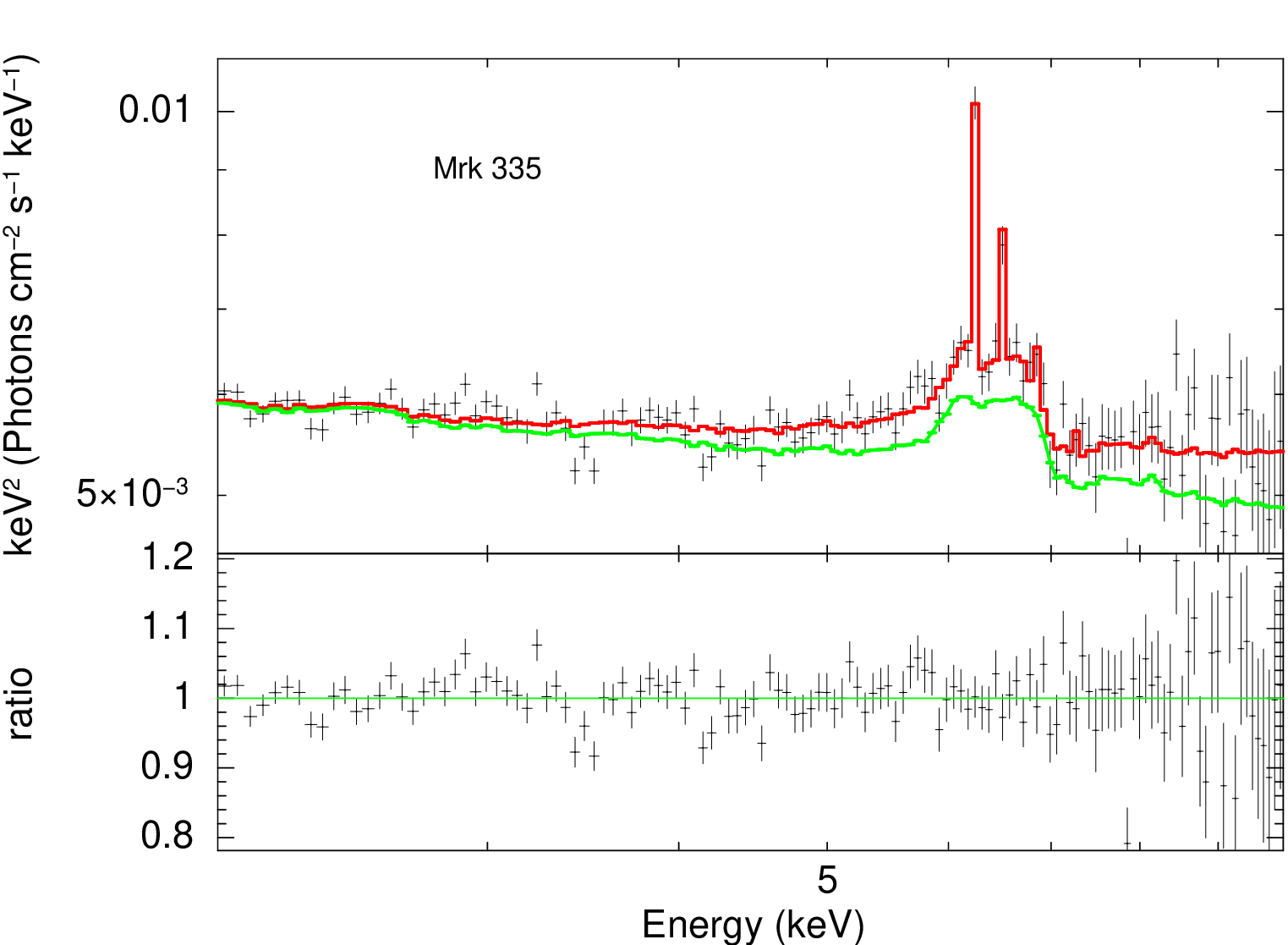}}
\rotatebox{-90}{\includegraphics[width=5cm]{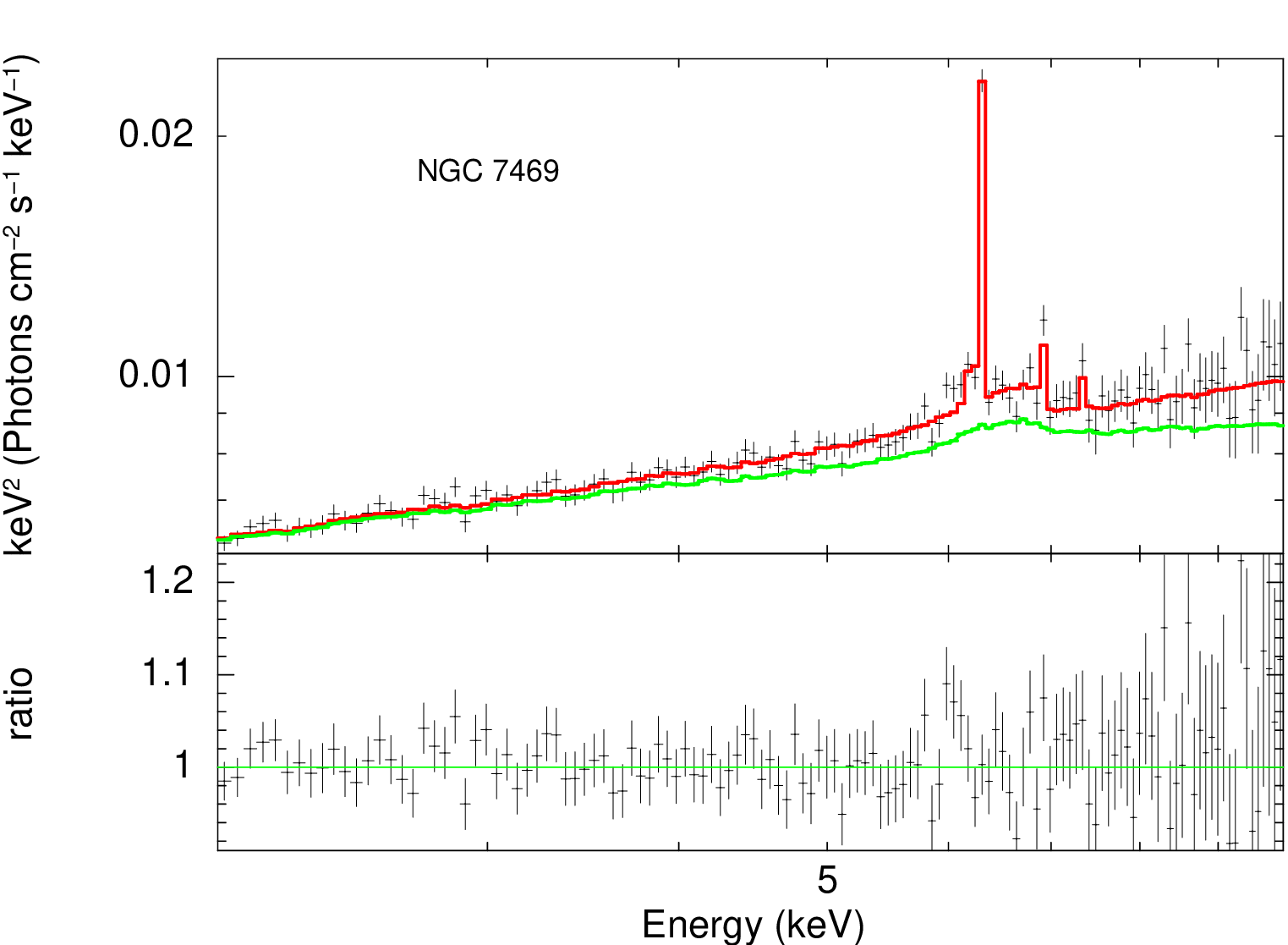}}\,\,\,\,\,\,\,\,\,\,\,\,\,\,\,\,\,\,\,\,\,\,\,\,\,\,\,\,\,\,\,\,\,\,\,\,\,\,\,\,\,\,\,\,
\rotatebox{-90}{\includegraphics[width=5cm]{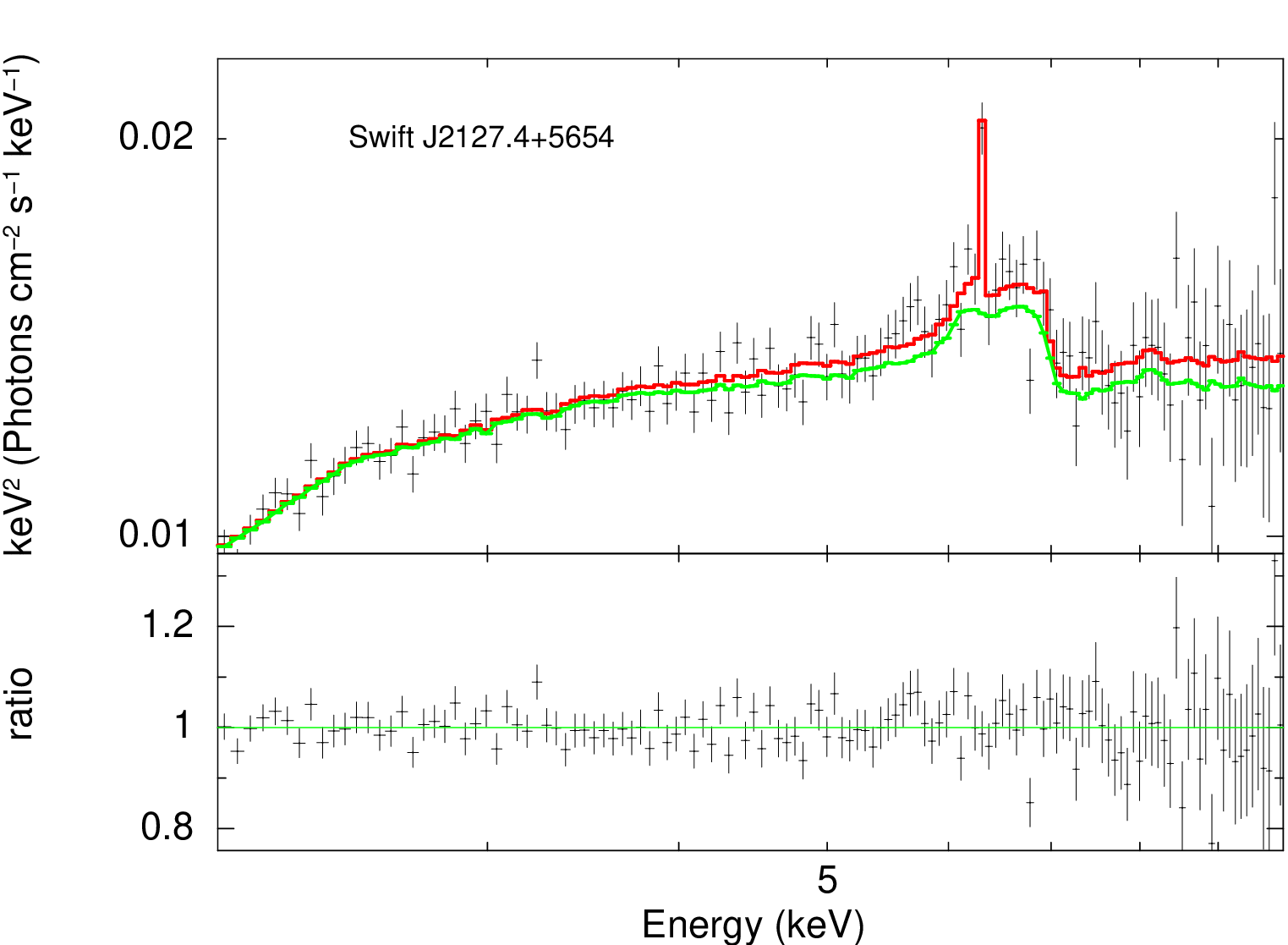}}
\caption{The 2-10 keV fits: The black markers represent the data, the green curve represents the wind model component, and the red curve represent the total spectrum. Two observations are shown overlaid for Fairall 9: for clarity, the second observation (OBSID 7050603010) has the data, wind model, and total spectrum represented by cyan, orange, and magenta, respectively.}
\label{fig:2_10}
\end{figure}

\begin{figure}
\rotatebox{-90}{\includegraphics[width=5cm]{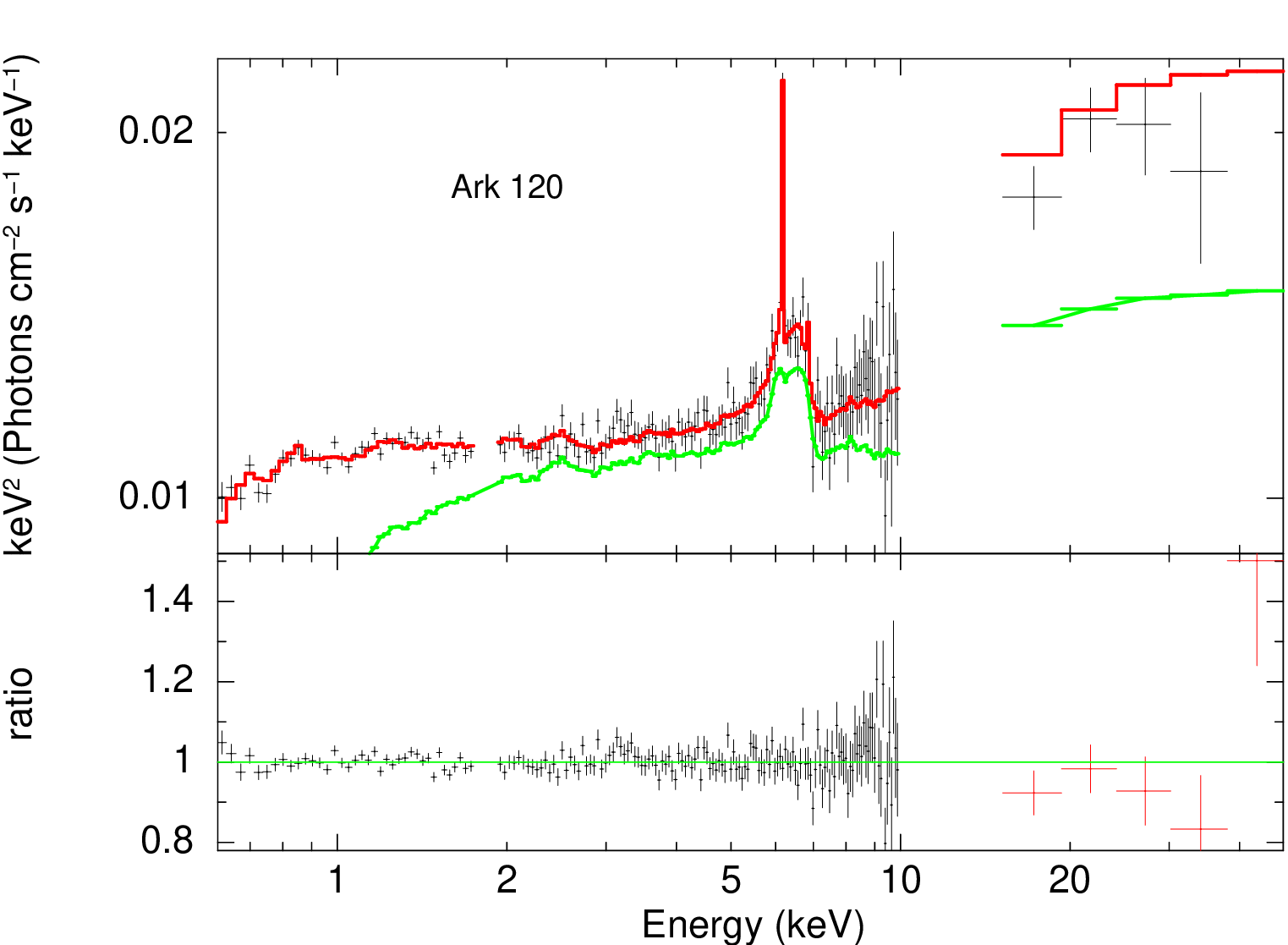}}
\rotatebox{-90}{\includegraphics[width=5cm]{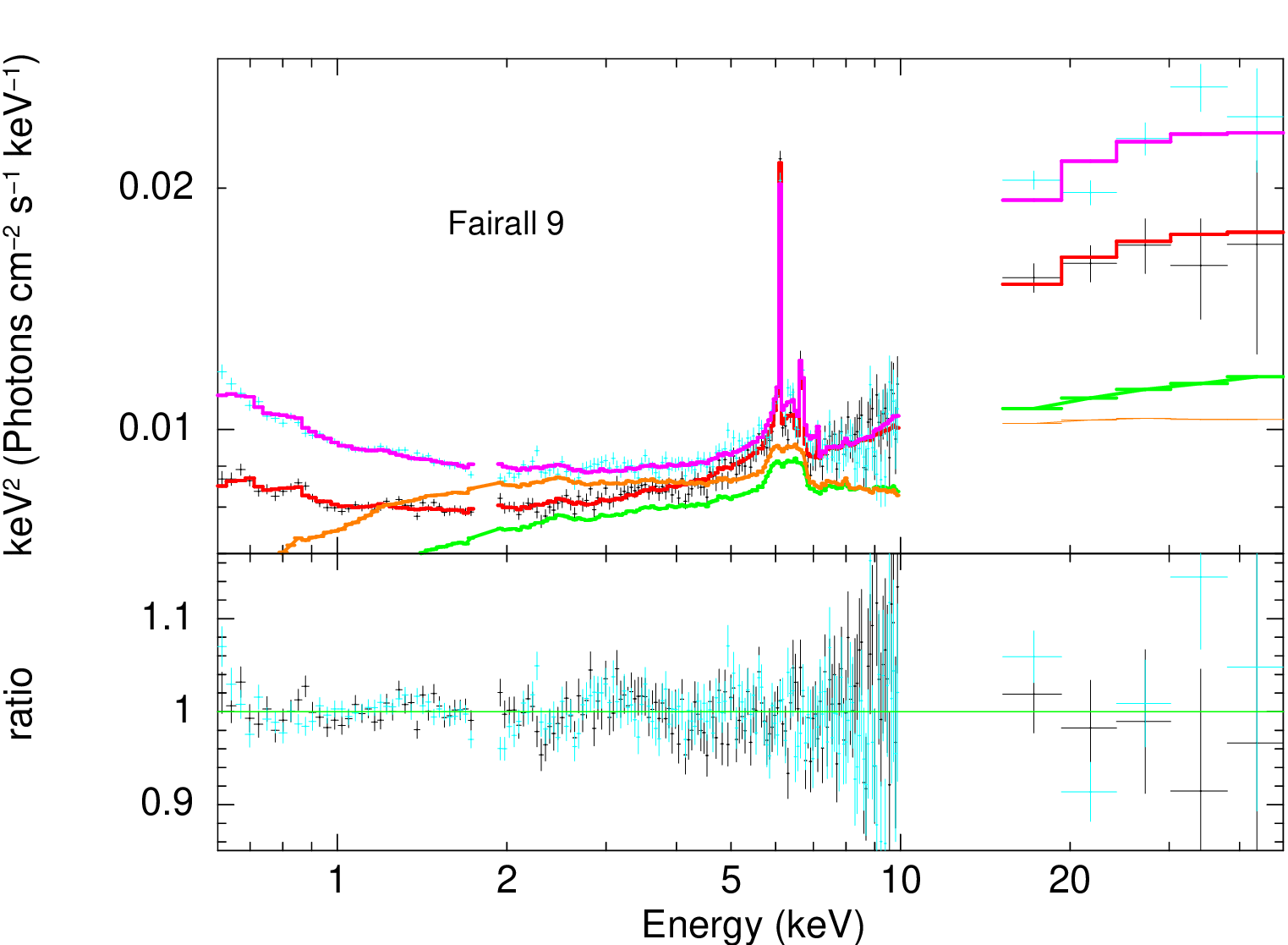}}
\rotatebox{-90}{\includegraphics[width=5cm]{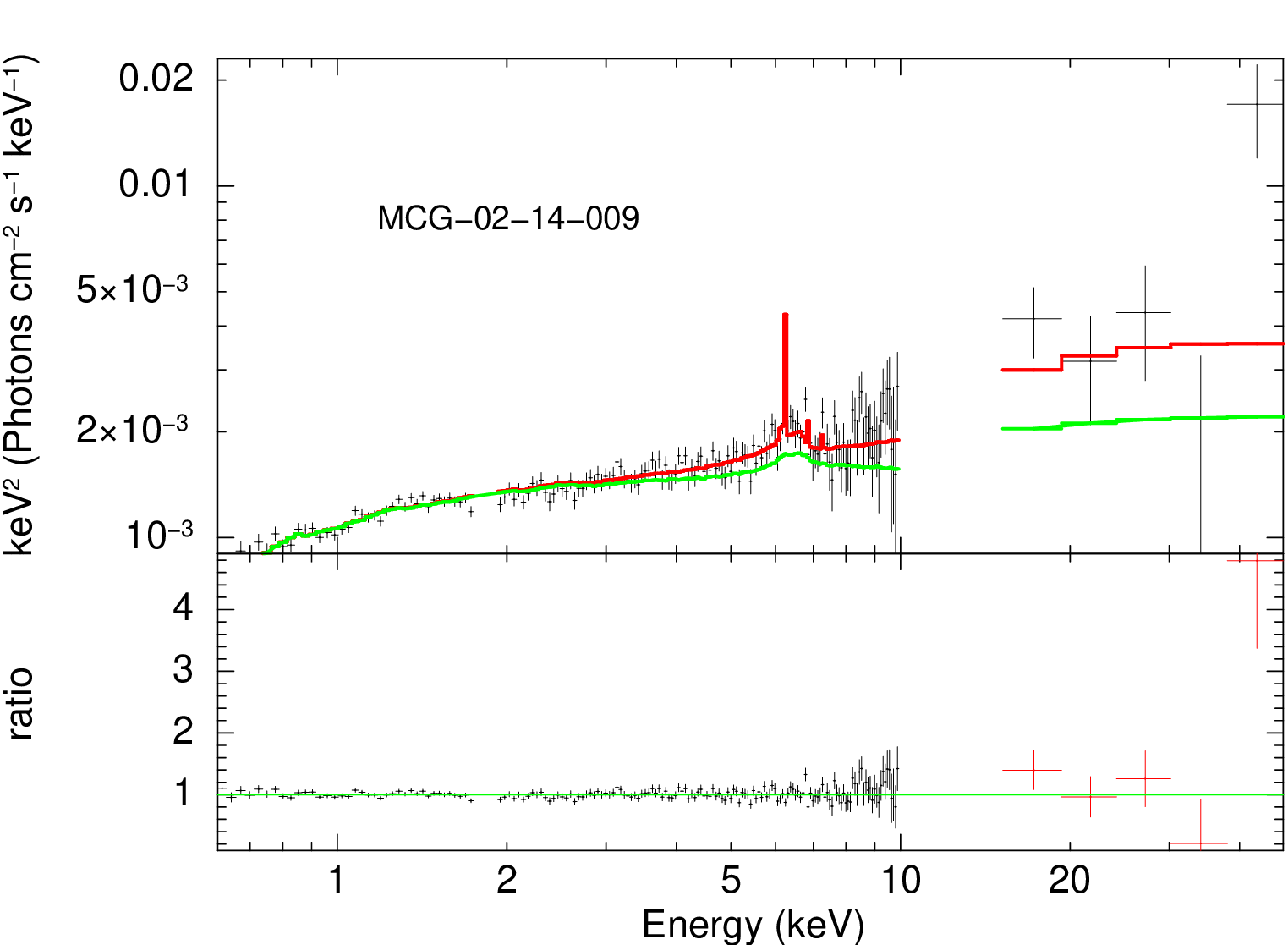}}
\rotatebox{-90}{\includegraphics[width=5cm]{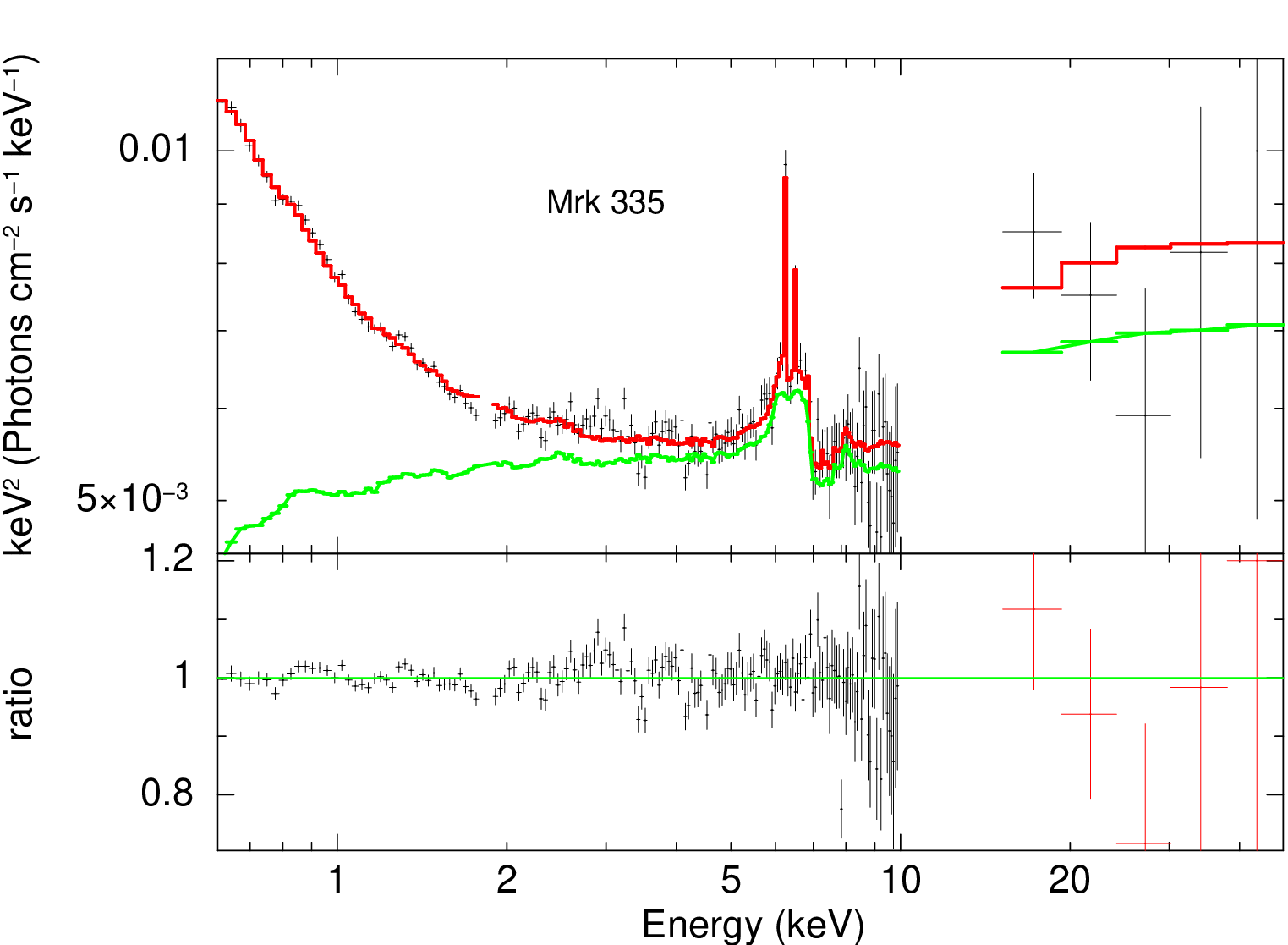}}
\rotatebox{-90}{\includegraphics[width=5cm]{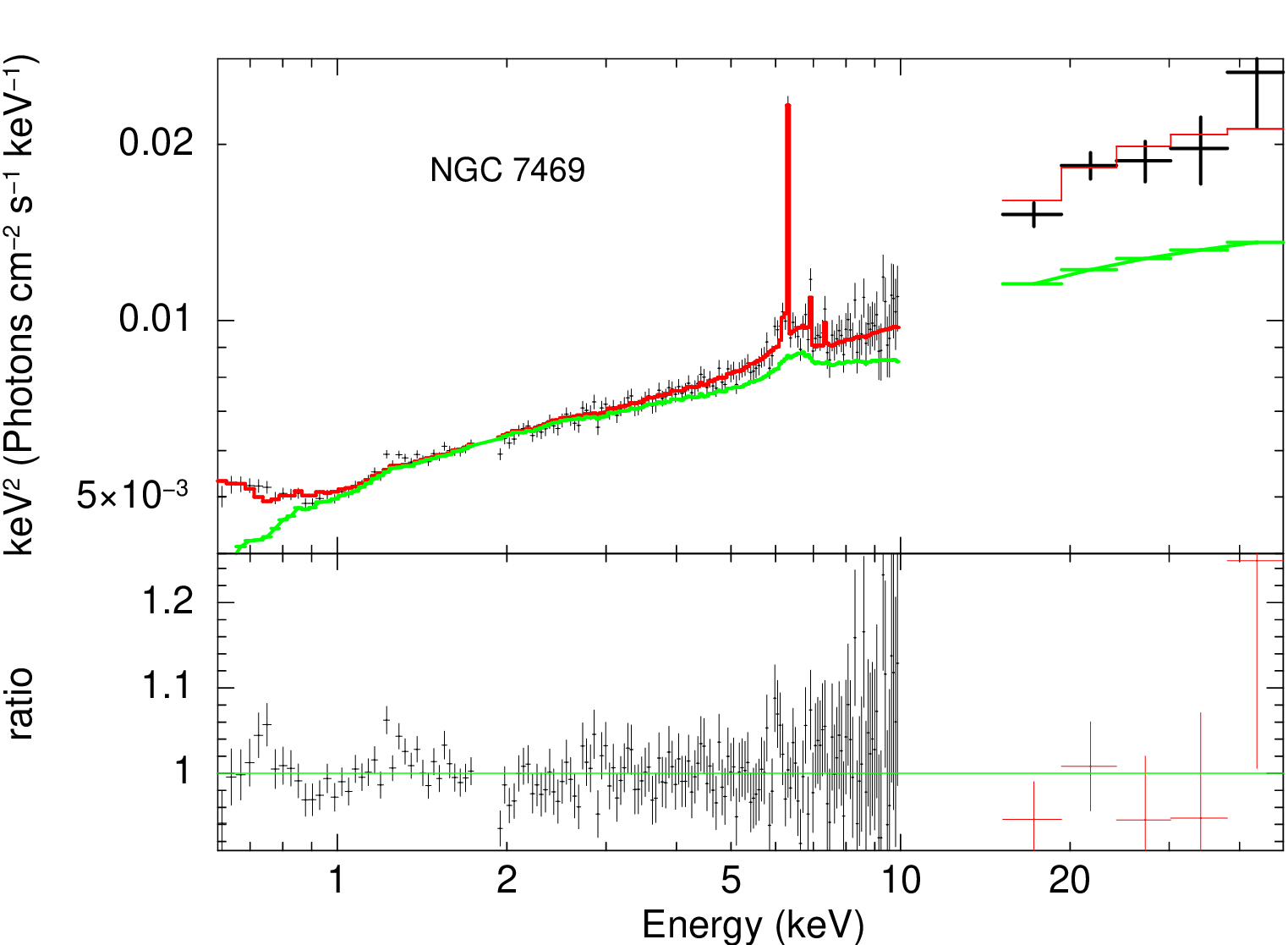}}\,\,\,\,\,\,\,\,\,\,\,\,\,\,\,\,\,\,\,\,\,\,\,\,\,\,\,\,\,\,\,\,\,\,\,\,\,\,\,\,\,\,\,\,
\rotatebox{-90}{\includegraphics[width=5cm]{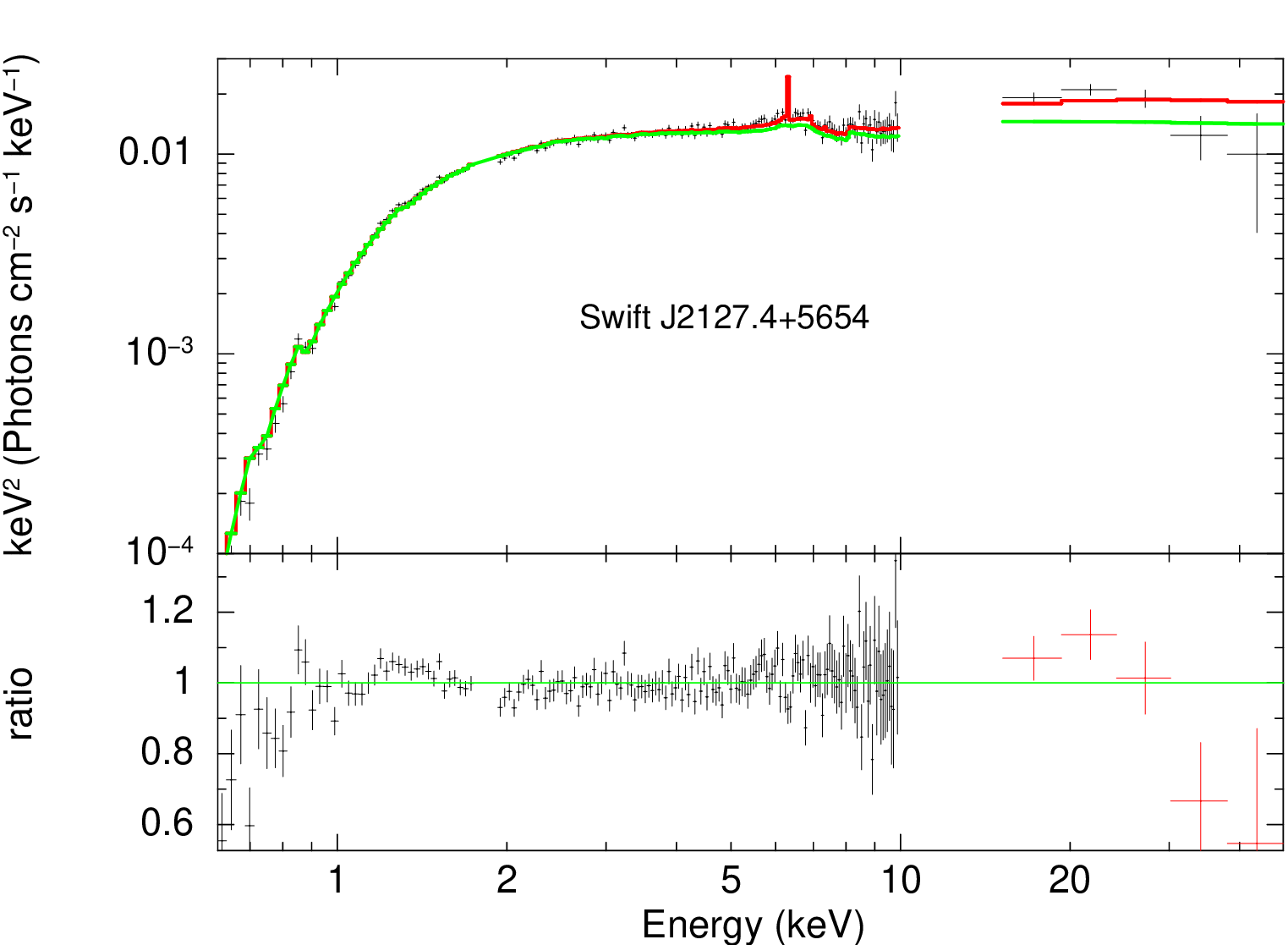}}
\caption{In the 0.6-50 keV fits: The black markers represent the data, the green curve represents the wind model component, and the red curve represent the total spectrum. Two observations are shown overlaid for Fairall 9: for clarity, the second observation (OBSID 7050603010) has the data, wind model, and total spectrum represented by cyan, orange, and magenta, respectively.}
\label{fig:All}
\end{figure}

Our initial model consisting of the wind table, a narrow Gaussian component and {\sc tbabs}  provided a good fit to the overall spectral curvature and, in particular,  to the moderately broad Fe K emission component of the sample sources. However, the model systemically underestimated the depth of the Fe K absorption edge. Adding the PIN data to the fit  and extrapolating the model up to 50 keV revealed that this construction also underestimated the PIN flux. Therefore, we considered whether adding a cold, neutral reflector would improve the fit. The addition of a cold, neutral reflector (fixed at R= --1) component to the total model (parameterized using {\sc pexmon}, \citealt{Nandra:2007lr}) produced an improvement to the fit to both the Fe K regime and the hard-band data (Figures~\ref{fig:2_10} and~\ref{fig:All}). {\sc pexmon} includes the Fe K$\alpha$, Fe K$\beta$ and Ni K$\alpha$ line emission expected from a reflecting slab, and no additional Gaussian component was required to explain the narrow core of Fe K emission. As a result, we removed the Gaussian component from the total model. Our final model construction for fits above 2 keV consisted of the wind model, {\sc tbabs}, and {\sc pexmon}.   

With the addition of the cold, neutral reflector,  the fits improved with $\Delta \chi^2$  $>$ 34 in 5/6 sources compared to fitting the 2-50 keV data without this component.  The reflector characterized by {\sc pexmon} constituted $\sim$ 10 -- 30 \% of the PIN flux, whereas the transmitted and scattered components of the wind constituted  $\sim$ 50 -- 85 \% of the PIN flux. The component contributions were measured against the total model predicted 15--50 keV band count rate. When statistically evaluating the merit of adding another component with the F-test, we found that the addition of the neutral reflector yielded an F-test probability of $<$ 10$^{-5}$ in 5/6 objects. This probability determines whether the null hypothesis can be rejected. Such a low F-test probability offers strong evidence for the addition of the extra component. We note that Ark 120 and Fairall 9 OBSID 705063010 required an additional component of Compton-thick gas (2 x 10$^{24}$ cm$^{-2}$) in the line-of-sight to fit the PIN data, with a 12\% and 31\% contribution to the PIN-band flux and a covering fraction of 8\% and 18\%, respectively.     

Despite the good statistical errors found for the wind model values, we note that the determination of errors is unlikely to be better than the coarseness of the model grid. The linear interpolation between grid points may yield solutions that do not accurately represent the true spectral shape of the model in those regimes, which in turn would yield incorrect wind model parameter values. 

To test the accuracy of solutions found using linear interpolation, we compared the model shape for one of the fits with a fine model grid calculated near the fit solution. The wind parameter values using this finer grid were within the 1$\sigma$ errors of the interpolated values with a comparable $\chi^{2}_{\nu}$. This suggests that our coarse model grid yields reasonable wind parameter values with a linear interpolation.

Because of the degeneracies in the model, the wind model parameter values must also be taken in context with the assumed geometry, i.e. the goal was to demonstrate that the Compton-thick wind model can explain the profiles of the broad components of Fe K$\alpha$ emission for a reasonable area of parameter-space.

\citet{Patrick:2011mz} identified residuals at $\sim 6.7$ and 6.9 keV compared to their baseline model  (Model A in their paper) . These features were attributed to narrow ($\sigma$=10 eV) ionized emission components from  Fe {\sc xxv} and Fe {\sc xxvi}, superposed on the broader spectral signatures. In our model, the scattering line from the wind contributes more flux in the 6.7-7.0 keV band than the \citet{Patrick:2011mz} model , giving a different baseline for any additional Gaussian line. From our analysis, inclusion of additional narrow Gaussian components in the model did not improve the fits in 4/6 objects, typically yielding  $\Delta \chi^2 \sim$ 1. However, the addition of a Gaussian component to the total model, statistically improved the fits ($\Delta\chi^2$ $>$ 10) in Mrk 335 and Fairall 9. A narrow ($\sigma$ =10 eV) Fe {\sc xxv} emission line at 6.67 $\pm$0.02 keV and a narrow ($\sigma$= 10 eV) Fe {\sc xxvi} emission line at 6.95 $\pm$ 0.05 keV were found in Mrk 335 and Fairall 9, respectively. The equivalent widths and line fluxes were 14$^{+8}_{-14}$ eV and 0.23$^{+0.13}_{-0.13}$ x 10$^{-5}$ photons cm$^{-2}$ s$^{-1}$ for Mrk 335 and 24$^{+6}_{-9}$ eV and 0.56$^{+0.16}_{-0.19}$ x 10$^{-5}$ photons cm$^{-2}$ s$^{-1}$ and 13$^{+7}_{-6}$ eV and 0.33$^{+0.27}_{-0.17}$ x 10$^{-5}$ photons cm$^{-2}$ s$^{-1}$ for Fairall 9 OBSID 702043010 and 705063010, respectively. The Mrk 335 and Fairall 9 OBSID 702043010 values were found to be similar to those tabulated in Model A of \citet{Patrick:2011mz}.   

Expanding the fits to 0.6-50 keV, we found a soft excess in 4 objects: Ark 120, Fairall 9, Mrk 335 and NGC 7469. In these objects, we parameterized the soft excess with the {\sc compTT} model which represents the Comptonization of soft photons in a hot plasma above the disk \citep{Titarchuk:1994fk}. A soft photon temperature of 0.02 keV was used to account for the soft excess. The plasma temperature and optical depth were found to be consistent with those values tabulated in Model A of \citet{Patrick:2011mz}. 

Fit parameters for the wind that were obtained over 0.6-50 keV are consistent with a majority of those derived from the  2-10 keV band (see Table~\ref{tab:Parameters}). This comparison demonstrates that the determination of the wind parameters is robust to the approach taken with fitting,  being based mainly upon the profile of the moderately broad Fe K-shell emission component. From Table~\ref{tab:Parameters}, the average photon index based on the 0.6-50 keV fits was $\Gamma=1.96^{+0.01}_{-0.01}$ and viewing angle $\theta=  59^{+4}_{-4}$ degrees. Excluding the poorly constrained mass-loss rate of Swift J2127.4+5654, the average mass-loss rate was 0.89$^{+0.25}_{-0.17}$ \.{M}$_{Edd}$, where \.{M}$_{Edd}$ is the Eddington mass-loss rate. The Eddington ratios for this sample (Table~\ref{tab:lum}) range from 0.01-0.03, which is consistent with the assumed L$_X$/ L$_{Edd}$ in the model. In the soft band of Mrk 335 and SWIFT J2127.4+5654, we note emission and absorption features in the residuals of the 0.6-50 keV fits. 
\section{Discussion}

\citet{Patrick:2011mz} identified this sample of six type I AGN having relatively simple X-ray spectra, with little evidence of any intrinsic X-ray absorption. Analysis of the sample sources showed them to possess both moderately broad and narrow components of Fe K-shell  emission. Modest broad components remaining after full consideration of the X-ray absorber complex have also been identified in NGC 5506 \citep{Guainazzi:2010yq} and NGC 3516   \citep{Markowitz:2008qy}. Further to these, moderately broad Fe K$\alpha$ emission components have been found on the variable component of the spectrum for  Mrk 766 \citep{Miller:2007fe} and NGC 4051 \citep{Miller:2010fj}. Thus, moderately broad (FWHM $\leq$ 1 keV) components of Fe K-shell emission appear quite common in the Seyfert population, and provide an important diagnostic of the circumnuclear gas. 

Using a relativistic diskline model to parameterize the broadness of the Fe K$\alpha$ emission component,  \citet{Patrick:2011mz} suggested that the lines originate from tens of gravitational radii, and that the profiles are not significantly modified by gravitational blurring.  

An alternative explanation for the moderately broad line component is that it arises from a Compton-thick  accretion disk wind. As shown by \citep{Sim:2008ys,Sim:2010fr}, reprocessing in a Compton-thick wind produces a strong broad line. In this case, rotational and outflowing velocities in the wind dominate the shape of the line profile, with an additional contribution from electron scattering. As is the case for the relativistic diskline model, the observed line shape can, in principle, provide information about the reprocessor, including the mass flow rate, physical state and geometry of the gas. When strong He- and H-like absorption signatures are detected, the wind model  provides an attractive link between the deep absorption lines from ionized species of Fe and the moderately broad Fe K-shell emission features \citep{Sim:2005uq,Sim:2008ys,Sim:2010fr}.  Prior to this work, comparisons between the Compton-thick wind and the observational data were  limited  to two case studies; the Fe K-shell emission and absorption in Mrk 766 and PG1211$+143$ have been shown to be consistent with arising in a wind \citep{Sim:2008ys,Sim:2010fr}.  Our study extends the comparison of the wind model with X-ray spectral data to an additional six sources. 

Application of the wind model results in  good  fits to the moderately broad component of Fe K-shell emission in all sample sources,  i.e. the breadth of the Fe K emission is consistent with the Compton scattering and velocity dispersion in our Compton-thick wind (Figures~\ref{fig:2_10},~\ref{fig:All}, Table~\ref{tab:Parameters}) of solar abundance, tested here across the radial range $\sim 30-300\, r_g$. Our conclusions are qualitatively similar to those of \citet{Patrick:2011mz} in that the moderately broad Fe K emission is consistent with an origin tens of r$_g$ from the black hole.  

Since Fe K-shell emission features are formed as a result of line scattering and recombination  in the outflow, their strength  depends on the density of material in the wind. In the model, the lines increase in strength with increasing \.{M}$_w$ \citep{Sim:2008ys,Sim:2010fr} and so, in principle, the observed line strength gives a way to measure the mass loss rate in the flow. In addition, better constraints can be placed on \.{M}$_w$ with higher spectral resolution in the Fe K regime.

The tabulated range for \.{M}$_w$ corresponds to  $\sim 2 - \,20$  times the accretion rate estimated for the sample AGN  \citep{Woo:2002vn,Petrucci:2004fr,Vaughan:2004rt, Miniutti:2009ys}, assuming a radiative efficiency of $\sim$0.1.  This means that our test has been limited to the case where outflow rates significantly exceed the accretion rate. However, such a test is reasonable, as  it is well established from UV data that mass outflow rates exceed the estimated accretion rate (e.g. \citealt{Crenshaw:2009fj}).  These fits  simply represent the case where most of the infalling material fails to reach the center of the AGN and accrete onto the black hole.  

A core component of Fe K$\alpha$ emission is present at 6.4 keV in all sources. The line core is too narrow to be produced in the wind, but  may arise via reflection from material further out. {\sc pexmon}, which links the narrow emission line strength to the reflected continuum, better constrained the contributions at 6.4 keV from a cold, neutral reflector. Using this parameterization, the calculated relative reflection values, R$_{frac}$, are poorly constrained, but range from 0.12 -- 4.48. In addition, we found a preferred inclination for the distant reflector to be 85$^o$ in most objects, based on the joint constraints offered by the line core and hard-band flux. Such an extreme solution suggests that {\sc pexmon} may not be the correct parameterization for the cold, neutral reflector. However, \citet{Yaqoob:2010fj} have shown that a toroidal X-ray reprocessor with Compton-thick lines-of-sight may have Fe K$\alpha$ line efficiencies that are orders of magnitude less than the maximum efficiency for that geometry, e.g. the Fe K$\alpha$ emission line strength/\,reflected continuum ratio may be different than those predicted by {\sc pexmon} allowing for weaker line strengths for a given hard-band flux. A less constrained line to reflected continuum ratio can, presumably, reduce the need for such an extreme inclination angle owing to having more reflection without significant production of Fe K$\alpha$.     

In addition to the narrow line core at 6.4 keV, \citet{Patrick:2011mz} reported the presence of narrow, ionized line emission from Fe {\sc xxv} and Fe {\sc xxvi} in some of the sample spectra. However, for the Compton-thick wind, the scattered Fe K line is predominantly formed as Fe {\sc xxv} and {\sc xxvi}, modified by scattering and velocity effects in the flow.  Thus, our wind model accounted for more flux in the  6.5-7.0 keV regime than the baseline model of \citet{Patrick:2011mz}, and we required no separate model components in that energy regime, with the exception of Fe {\sc xxv} emission and Fe {\sc xxvi} emission in Mrk 335 and Fairall 9, respectively. 

There is little evidence for  Fe {\sc xxv} and Fe {\sc xxvi} absorption in the observed spectra of the sample. Spectral fit results suggest intermediate viewing angles for most of these wind systems, with values clustered about the mean angle $\theta=  59^{+7}_{-7}$ degrees. Systems at such orientations are expected to present  a strong contribution from the scattered component of the wind  \citep{Sim:2008ys,Sim:2010fr}, and this, in part, explains the lack of strong predicted absorption lines in the total spectrum. 

Further to this, as noted previously, several parameters were fixed due to degeneracies in the model and the values found for $\theta$ and \.{M}$_w$ must be taken within the context of the assumed geometry. In comparison to the model parameters assumed in \citet{Sim:2010fr}, our geometry placed the inner edge of the wind at a relatively small inner radius and large outer radius, resulting in a comparatively large wind opening angle and  making the mass distribution of the wind  significantly different to that employed by \citet{Sim:2010fr}.  Our  chosen geometry resulted in a relatively high ionization parameter compared to the model discussed by \citet{Sim:2010fr}, owing to a reduced mean gas density  and also less geometric dilution of the ionizing flux at the inner edge of the wind. Our chosen parameters also resulted in a faster flow. The higher ionization state of the gas results in a lower fraction of Fe {\sc xxv} and Fe {\sc xxvi} ions and so reduces the depth of the absorption features predicted by our solutions, compared to those predicted by \citet{Sim:2010fr}. The faster velocity of our flow also broadens the predicted absorption features, making them less prominent in the model spectra. To achieve absorption features of a similar depth to those predicted by \citet{Sim:2010fr}, we would need to view the wind model presented here at higher angles than were found in the spectral fitting. In summary, the depth of Fe {\sc xxv} and Fe {\sc xxvi} absorption lines in the model spectra depend on the wind geometry, mass outflow rate and  viewing angle, and there is no inconsistency between model predictions and the inclination angles found in our fits.

Future observations with higher spectral resolution and signal-to- noise may be able to resolve the subtle spectral features in the model and allow all wind parameters to be determinable. In addition, soft band emission lines, such as O K$\alpha$, Si K$\alpha$ and S K$\alpha$, may help to constrain some of the wind parameters. 

\section{Conclusions}
We have taken a sample of sources that show moderately broad components of Fe K-shell emission, and fit them using a model for a Compton-thick outflow. All moderately broad Fe K emission components are well described by the wind model,  demonstrating the general applicability of such models to the interpretation of X-ray emission lines. Such an approach complements studies of X-ray winds based on X-ray  absorption features. Solutions found for the sample sources also show that 50 -- 85 \% of the flux in the 15 -- 50 keV band has been transmitted through or scattered by the wind.

We conclude that the X-ray signatures of a Compton-thick wind have a profound effect on the observed X-ray spectra of AGN, even when the nuclear system is not viewed through the wind. In principle, fits to X-ray spectra of AGN can provide important constraints on the wind mass loss rate, geometry and orientation of the nuclear wind. Future work will extend this study to more  complex sources, likely viewed at somewhat higher inclination angles.  

\section{Acknowledgements}
MMT would like to acknowledge NASA grant NNX10AL83H. TJT would like to acknowledge NASA grant NNH10ZDA001N-ADAP. LM acknowledges support from STFC grant  ST/H002456/1, which also funded the wind model computation carried out at Oxford University Department of Physics.

\bibliographystyle{apj}      
\bibliography{list} 

\newpage
\begin{deluxetable}{lccc}
\tabletypesize{\scriptsize}
\tablecaption{Parameters for Model Grid} 
\tablewidth{0pt}
\startdata
\hline
\hline
Parameter&&Values&\\
&&&\\
\hline
&&&\\ 
log$_{10}$ \.{M}$_w$/\.{M}$_{Edd}$&-0.5& -0.1& 0.4\\
log$_{10}$ R$_{min}$/(100 R$_g$)& -0.5&  0.0&  0.5 \\
log$_{10}$ {\it d}/R$_{min}$ &0.0&  0.5& \\
R$_{max}$/R$_{min}$& 3.0&  9.0&\\
Fe abundance/solar &1.0&  3.0 &\\
$\Gamma$ &1.8& 2.4& 3.0\\
$f_v$&1&&\\
$R_v$&$R_{max}$&&\\
$\beta$&1&&\\
$v_0$&0&&\\
M/M$_{\odot}$&10$^7$&&\\
$\kappa$&--1&&\\
L$_X$/L$_{Edd}$&0.01&&\\
\enddata 
\tablecomments{The parameters with multiple values are the wind model grid points, while the parameters with a single value are assumed in the wind model. See section 2.1 for details.}
\label{tab:syn}
\end{deluxetable}

\begin{deluxetable}{lllcccccccc}
\tabletypesize{\scriptsize}
\tablecaption{Wind Fits} 
\tablewidth{0pt}
\tablehead{
\colhead{Object} &
\colhead{$\Gamma$} & 
\colhead{\.{M}$_w$ \tablenotemark{1} }  & 
\colhead{$\theta$ \tablenotemark{2}} & 
\colhead{$N_r$ \tablenotemark{3}} & 
\colhead{$\theta$$_r$ \tablenotemark{4}} & 
\colhead{ kT\tablenotemark{5}}&
\colhead{$\tau$\tablenotemark{6}}&
\colhead{$R_{frac}$\tablenotemark{7}}&
\colhead{$N_H$    \tablenotemark{8}} &
\colhead{$\chi^2_\nu /dof$}\\
}
\startdata
\multicolumn{11}{c}{\textbf{2-10 keV}}\\ \\
Ark 120 & 2.02$^{+0.02}_{-0.03}$ & $0.84^{+0.21.}_{-0.14}$ & 63$^{+5}_{-2}$&22.2$^{+5.8}_{-14.2}$ & 85$^{\it +0}_{-45}$&--&--&1.00$^{+0.48}_{-0.72}$& 9.8$^{+3.5}_{\it -0}$ & 1.05/124\\  \\
Fairall 9 (702)\tablenotemark{9} & 1.95$^{+0.02}_{-0.04}$&0.53$^{+0.85}_{\it -0.20}$&38$^{+12}_{-7}$&25.2$^{+1.2}_{-13.3}$&$85^{\it +0}_{-4}$&--&--&3.89$^{+0.50}_{-3.14}$&3.2$^{+7.8}_{\it -0}$&1.23/250\\ \\
Fairall 9 (705)\tablenotemark{9} &2.02$^{+0.01}_{-0.03}$&0.78$^{+1.10}_{-0.25}$&$''$&25.8$^{+3.2}_{-14.8}$&$''$$^{}_{}$&--&--&3.45$^{+0.47}_{-3.04}$&$''$&--\\ \\
MCG-02-14-009&1.84$^{+0.08}_{\it -0.04}$&0.82$^{\it +1.77}_{\it -0.49}$&$29^{+20}_{\it -11}$&3.2$^{+0.8}_{-2.8}$&85$^{\it +0}_{\it -85}$&--&--&3.37$^{+1.19}_{-2.99}$&10.7$^{+24.7}_{\it -1.5}$&1.02/124\\ \\
Mrk 335\tablenotemark{9}&2.11$^{+0.02}_{-0.02}$&0.54$^{+0.16}_{-0.10}$&65$^{+2}_{-5}$&12.5$^{+3.5}_{-9.5}$&85$^{\it +0}_{\it -85}$&--&--&1.05$^{+0.13}_{-0.82}$&$3.6^{+6.7}_{\it -0}$&1.62/122\\ \\
NGC 7469&1.80$^{+0.04}_{\it -0 }$&0.87$^{\it +1.69}_{\it -0.54 }$&46$^{+19}_{-18}$&$3.9^{+17.5}_{-2.47}$&38$^{\it +47}_{\it -38}$&--&--&0.80$^{+4.20}_{-0.64}$&9.7$^{+8.3}_{\it -5.2}$&0.76/124\\ \\
SWIFT J2127.4&2.02$^{+0.03}_{-0.04}$&0.46$^{+0.30}_{\it -0.13}$&69$^{+2}_{-9}$&13.4$^{+3.4}_{-11.6}$&85$^{\it +0}_{\it -85}$&--&--&0.40$^{+0.36}_{\it -0.40}$&$79.1^{+ 11.2}_{\it -2.6}$&0.97/124\\
\\ 
\multicolumn{11}{c}{\textbf{0.6-50 keV\tablenotemark{9}}}
\\ \\
Ark 120& $1.97^{+0.03}_{-0.03}$ & 0.96$^{+0.10}_{-0.21}$ & 64$^{+7}_{-3}$& 17.7$^{+4.3}_{-13.7}$&85$^{\it +0}_{-30} $ &$<$ 11&$<$ 1.0&0.78$^{+0.35}_{-0.66}$&10.2$^{+6.8}_{\it -0.4}$& 1.26/166\\ \\
Fairall 9 (702)& 1.93$^{+0.02}_{-0.03}$&0.55$^{+0.20}_{- 0.15}$ &65$^{+6}_{-1}$&$23.1^{+4.4}_{-14.1}$&$85^{\it +0}_{-10}$&$<$ 8.7&$<$ 2.5&1.68$^{+1.85}_{\it -1.68}$&7.4$^{+4.6}_{-2.4}$&1.34/332\\ \\
Fairall 9 (705)&2.02$^{+0.02}_{-0.03}$&0.52$^{+0.13}_{-0.09}$&$''$&$24.5^{+5.5}_{-13.9}$&$''$$^{}_{}$&$<$ 8.5&$<$ 2.3&1.51$^{+1.49}_{-1.12}$&$''$&--\\ \\
MCG-02-14-009&1.90$^{+0.03}_{-0.02}$&1.71$^{\it +0.85}_{-0.82}$&43$^{+19}_{-15}$&4.6$^{+1.4}_{-1.9}$&85$^{\it +0}_{-3}$&--&--&4.48$^{+1.92}_{-3.57}$&9.2$^{+0.5}_{\it -0}$&1.20/170\\ \\
Mrk 335&2.00$^{+0.03}_{-0.02}$&0.51$^{+0.17}_{-0.10}$&71$^{+2}_{-5}$&$1.9^{+0.8}_{-0.9}$&$63^{\it +22}_{\it -63}$&$<$ 13.5&$<$ 2.0&0.12$^{+0.08}_{-0.11}$&3.7$^{+3.3}_{\it -0.1}$&1.88/166\\ \\
NGC 7469&1.85$^{+0.02}_{-0.01}$&0.86$^{+1.19}_{-0.56}$\ &44$^{+10}_{-6}$&2.3$^{+0.8}_{-0.7}$ &85$^{\it +0}_{-10}$&$<$ 13.9&$<$ 1.2&0.46$^{+0.24}_{-0.28}$&8.6$^{+4.4}_{-1.6}$&1.05/166\\ \\
SWIFT J2127.4&2.07$^{+0.02}_{-0.02}$&$0.33^{+0.40}_{\it -0}$&69$^{+2}_{-12}$&26.7$^{+6.3}_{-6.7}$&85$^{\it +0}_{-3}$&--&--&0.83$^{+0.18}_{-0.23}$&$79.0^{+1.4}_{-1.3}$&1.55/169\\
\enddata
\tablecomments{Errors are calculated at 90\% confidence. Values in italics represent ranges truncated by the limit imposed during the fit.} 
\tablenotetext{1}{The flow rate in \.{M}$_{Edd}$, the Eddington mass-loss rate}
\tablenotetext{2}{The observer's viewing angle through the wind, in degrees}
\tablenotetext{3}{Normalization of the {\sc pexmon} component in units 10$^{-3}$ photons cm$^{-2}  {\rm s^{-1}}$}
\tablenotetext{4}{The angle of inclination of the {\sc pexmon} component, in degrees}
\tablenotetext{5} {{\sc compTT} Plasma temperature, in keV} 
\tablenotetext{6} {{\sc compTT} Plasma optical depth}
\tablenotetext{7}{The calculated reflection scaling factor between {\sc pexmon} and the intrinsic powerlaw continuum }
\tablenotetext{8}{Galactic column density in units of 10$^{20} {\rm atom\, cm^{-2}}$, with the values from the DL HI in the Galaxy survey as a lower limit}
\tablenotetext{9}{See text for details}

\label{tab:Parameters}

\end{deluxetable}

\newpage
\begin{deluxetable}{lccc}
\tabletypesize{\scriptsize}
\tablecaption{2-10 keV Luminosity} 
\tablewidth{0pt}
\tablehead{
\colhead{Object} &
\colhead{L$_{obs}$\tablenotemark{1}}&
\colhead{L$_X$\tablenotemark{2}}&
\colhead{L$_X$/L$_{Edd}$}\\
}
\startdata
\multicolumn{4}{c}{}\\
Ark 120 \tablenotemark{3}&7.5E43&2.1E44&0.01\\
Fairall 9 \tablenotemark{3} (702043010)&1.2E44&2.8E44&0.03\\
Fairall 9 \tablenotemark{3} (705063010)&1.3E44&2.6E44&0.03\\
MCG-02-14-009 \tablenotemark{4}&7.8E42&1.6E43&--\\
Mrk 335 \tablenotemark{5}&2.3E43&5.9E43&0.02\\
NGC 7469 \tablenotemark{3}&1.3E43&2.5E43&0.03\\
SWIFT J2127.4+5654 \tablenotemark{6}&1.6E43&4.6E43&0.03\\
\enddata 
\tablenotetext{1}{The observed 2-10 keV luminosity in erg s$^{-1}$}
\tablenotetext{2}{The intrinsic 2-10 keV luminosity in erg s$^{-1}$}
\tablenotetext{3}{Black hole mass estimate from \citet{Woo:2002vn}}
\tablenotetext{4}{Black hole mass estimate not found in the literature}
\tablenotetext{5}{Black hole mass estimate from  \citet{Grier:2012vn}}
\tablenotetext{6}{Black hole mass estimate from \cite{Malizia:2008yq}} 
\label{tab:lum}
\end{deluxetable}

\end{document}